# GYROKINETIC EQUATIONS FOR STRONG-GRADIENT REGIONS

Andris M. Dimits[a)]

*Lawrence Livermore National Laboratory, Livermore, California 94550, USA*

A gyrokinetic theory is developed under a set of orderings applicable to the edge region of tokamaks and other magnetic confinement devices, as well as to internal transport barriers. The result is a practical set equations that is valid for large perturbation amplitudes [$q\delta\psi/T = O(1)$, where $\delta\psi = \delta\phi - v_\parallel \delta A_\parallel/c$], which is straightforward to implement numerically, and which has straightforward expressions for its conservation properties. Here, $q$ is the particle charge, $\delta\phi$ and $\delta A_\parallel$ are the perturbed electrostatic and parallel magnetic potentials, $v_\parallel$ is the particle velocity, c is the speed of light, and T is the temperature. The derivation is based on the quantity $\varepsilon \equiv (\rho/\lambda_\perp) q\delta\psi/T \ll 1$ as the small expansion parameter, where $\rho$ is the gyroradius and $\lambda_\perp$ is the perpendicular wavelength. Physically, this ordering requires that the $E \times B$ velocity and the component of the parallel velocity perpendicular to the equilibrium magnetic field are small compared to the thermal velocity. For nonlinear fluctuations saturated at "mixing-length" levels (i.e., at a level such that driving gradients in profile quantities are locally flattened), $\varepsilon$ is of order $\rho/L_p$, where $L_p$ is the equilibrium profile scale length, for all scales $\lambda_\perp$ ranging from $\rho$ to $L_p$. This is true even though $q\delta\psi/T = O(1)$ for $\lambda_\perp \sim L_p$. Significant additional simplifications result from ordering $L_p/L_B = O(\varepsilon)$, where $L_B$ is the spatial scale of variation of the magnetic field. We argue that these orderings are well satisfied in strong-gradient regions, such as edge and screapeoff layer regions and internal transport barriers in tokamaks, and anticipate that our equations will be useful as a basis for simulation models for these regions.

---

[a)]Email address: dimits1@llnl.gov

## I. Introduction

Strong ambient magnetic fields can result in plasma confinement over time scales many orders of magnitude longer than the characteristic thermal-escape or free-expansion time that would result in the absence of the magnetic field, given the plasma temperature and spatial scale.[1] This is a key feature of space and astrophysical plasmas (for example, the magnetospheres of stars and planets) and is the basis for magnetically confined laboratory plasmas, which constitute one approach to controlled nuclear fusion energy ("magnetic fusion energy" or "MFE"). The dominant (most rapid) charged particle motion in strong magnetic fields is the approximately circular cyclotron or gyro-orbital motion of the particles. Both the remaining single-particle motions and many collective phenomena of interest occur on time scales that are much longer than the gyro-orbit frequency (cyclotron or gyrofrequency). For example, turbulence driven by instabilities with frequencies well below the ion cyclotron frequency and wavelengths of order or longer than the ion gyroradius is believed to be a key cause of anomalous transport of heat, momentum and particles that is observed in magnetic fusion plasmas. In such situations, the magnetic moment becomes an adiabatic invariant of the particle motion.[1] The magnetic moment can be written to lowest order as $\mu = v_\perp^2/2\Omega$, where $v_\perp$ is the magnitude of the velocity perpendicular to the magnetic field, $\Omega = qB/Mc$ is the cyclotron frequency, $q$ is the particle charge, $B$ is the magnetic field strength, $M$ is the particle mass, and $c$ is the speed of light. The fundamental requirement for the magnetic moment of a charged particle in a magnetic field to be an adiabatic invariant can be written as

$$\varepsilon = \frac{\omega}{\Omega} \ll 1, \quad (1)$$

where $\omega$ is a measure of the relative rate of change of the electromagnetic fields seen by the particle.

The adiabatic invariance of the magnetic moment has been exploited to simplify the description of the particle motion and collective plasma behavior through the guiding-center[1,2] and gyrokinetic theories.[3-11] The guiding-center and gyrokinetic equations both separate the particle motion into the orbital motion, which is averaged over so that the equations for the other variables do not depend on the orbital phase, as well as the parallel motion and perpendicular drift motions of the orbital centers ("guiding centers" and "gyro centers" respectively in the guiding-center and gyrokinetic theories) relative to the magnetic field lines. This separation makes these systems easier and more intuitive to use for investigations of plasma phenomena in situations where the guiding-center and gyrokinetic theories are applicable. The guiding center theory in particular has played a key role in the understanding the most fundamental aspects of particle confinement in magnetized plasmas in space and laboratory settings.[1] The theories differ in that the guiding-center theory assumes that the particle gyro-orbit size is small compared to the spatial scales of any inhomogeneities in the system, while the gyrokinetic theory allows for variations of the electric and magnetic fields on spatial scales comparable to the gyroradius, provided that these variations are sufficiently small according to a "gyrokinetic ordering," which will be discussed in more detail shortly. This aspect makes the gyrokinetic equations more relevant for studies of plasma microinstabilities, for which often a significant part of the spectrum of interest has spatial scales comparable to the gyroradius. The gyrokinetic equations have also found wide use in numerical simulations,[12-20] largely because they reduce the dimensionality of the relevant phase space by one relative to the raw "full-dynamics" equations and, in many settings, are also less computationally stiff.[12]

The methods used to derive the gyrokinetic equations can be crudely grouped into Hamiltonian[7-9,11] and non-Hamiltonian methods.[3-6,10] Non-Hamiltonian derivations of the gyrokinetic equations, which typically use multiscale perturbation theory applied directly to the Vlasov equation and/or its characteristics (the equations of motion), predate the Hamiltonian derivations. The conservation laws at each order in the non-Hamiltonian theories are typically only approximately satisfied, with higher order terms acting as sources or sinks in the conservation equations.

The Hamiltonian derivations[7-9,11] are largely based on the prior related work on guiding-center theory by Littlejohn,[2] and a clear exposition of the powerful noncanonical version of this method, along with the application to the derivation of the drift-kinetic equations for a particle in an inhomogeneous magnetic field, was given in Ref. 2. In the Hamiltonian derivations, the perturbation theory is carried out directly on the particle Lagrangian or Hamiltonian (in the non-canonical and canonical versions, respectively). The resulting Lagrangian can be incorporated into a system action integral from which, with the use of Noether's theorem, it follows that at each order of the theory there are versions of the physical conservation laws (e.g., energy, momentum, angular momentum) that are satisfied exactly.[9,21] This is a valuable attribute of the Hamiltonian theories, and a clear difference with the results of the non-Hamiltonian derivations. The noncanonical Hamiltonian method was applied to the derivation of the gyrokinetic particle (characteristic), Vlasov, and Poisson equations for a plasma in a toroidal magnetic field by Hahm.[8] Subsequently, generalizations were made for a variety of situations, and this body of work has been reviewed by Brizard and Hahm.[9]

Most Hamiltonian nonlinear gyrokinetic theories[7-9,11] are based on what we will call here the "standard gyrokinetic ordering," which takes the form

$$\frac{q\delta\psi}{T} \sim \frac{\delta f}{F_{eq}} \sim k_\| \rho_t \sim \varepsilon \ll 1, \quad (2)$$

Here, $\delta\psi = \delta\phi - v_\| \delta A_\| / c$, $\delta\phi = \delta\phi(\boldsymbol{x},t)$ is the perturbed electrostatic potential, $\delta A_\|$ is the perturbed parallel magnetic potential, $v_\|$ is the parallel particle velocity, c is the speed of light, $T$ is the temperature, $\delta f$ and $F_{eq}$ are the perturbed and equilibrium phase-space distribution functions, $k_\|$ is the characteristic parallel wavenumber, $\rho_t = v_t/\Omega$ is the thermal gyroradius, $v_t = \sqrt{T/M}$ is the thermal (kinetic) particle speed, and $\varepsilon$ is the formal expansion parameter ("gyrokinetic smallness parameter"). This statement of the ordering allows for electromagnetic perturbations, and assumes an additional low-$\beta$ (where $\beta$ is the ratio of the plasma to magnetic pressure) or "shear-Alfven" ordering for the perturbed magnetic vector potential. The first condition of Eq. (2) is somewhat counterintuitive as stated, because it suggests that the theory would break down if a large constant were added to $\delta\psi$, and makes more sense if $\delta\psi$ is replaced by some measure of its spatial variation. A partial resolution of this apparent paradox lies in that these theories also formally order $k_\perp \rho_t \sim 1$, where $k_\perp$ is the characteristic perpendicular wavenumber. Equation (2) is consistent in most cases (absent very large-scale instabilities) with tokamak core plasma conditions. Indeed the majority of magnetic fusion relevant gyrokinetic simulation codes and investigations reported to date have been for MFE core plasmas,[12-20] and are based on this ordering.

An additional ordering used in most versions of the gyrokinetic theory that allow for spatially inhomogeneous (e.g., toroidal) equilibrium magnetic fields[8-11] is that the spatial scale of variation of the magnetic field $L_B$, which for toroidal systems is typically of order the major radius $R$, is ordered to be the same as the spatial scale $L_p$ over which the profile quantities that drive the instabilities and turbulence vary, i.e.,

$$\frac{L_p}{L_B} = O(1). \qquad (3)$$

It then follows that if $\|\boldsymbol{\nabla} B\|$ is a measure of the spatial derivatives of the magnetic field $\boldsymbol{B}$, then

$$\frac{\rho_t \|\boldsymbol{\nabla} B\|}{B} = O(\varepsilon). \qquad (4)$$

In the majority of such work,[8,9] the theory is developed to second or higher order in the perturbed quantities, but only to the first order (the order at which the magnetic drifts first appear) in the equilibrium magnetic field inhomogeneities. This choice enables the use of a "two-step" derivation in which the steps are as follows. The guiding-center Lagrangian for the particle in the inhomogeneous magnetic field is first calculated using a perturbative transformation (to first order in $\varepsilon$) to remove the gyrophase dependences associated with the equilibrium magnetic-field terms. Then, a second perturbative transformation is used to remove the gyrophase dependences from the perturbed Lagrangian (which is the sum of the equilibrium guiding-center Lagrangian and the Lagrangian for the perturbations) to second or higher order in $\varepsilon$. This procedure has been preferred largely because it is somewhat simpler than the more direct "one-step" procedure in which a single perturbative transformation is used to simultaneously remove the gyrophase dependences from both the equilibrium and perturbation terms in the Lagrangian, and it most easily builds upon prior guiding-center theory results. However, the application of two-step procedure in the manner of the aforementioned work does not result in a Lagrangian that is consistent to second order in $\varepsilon$ under the orderings of Eqs. (2)-(4). Using a one-step procedure, a systematic derivation of the gyrokinetic theory, including the Lagrangian to second and higher orders, has recently been carried out for the first time.[11] Even at second order, many additional terms result beyond those given, for example, in Refs. 8 and 9, and these terms pose significant challenges for analytical work or simulations that attempt to go beyond first order in $\varepsilon$.

There is now considerable interest in applying gyrokinetic models to strong-gradient regions in tokamaks and other magnetic plasma confinement devices, for example the outer edge and scrapeoff-layer regions,[22,23] and internal transport barriers.[24] The existence of large potential and density fluctuations in the edge and scrapeoff layer regions of tokamaks, and large quasi-steady potentials that violate the standard gyrokinetic ordering of Eq. (2) has long been known,[25,26] and these have also been observed in fluid simulations of the edge and scrapeoff-layer regions.[27] In addition, large electrostatic potentials that are inconsistent with the standard ordering have been inferred from observations of flows in internal transport barriers.[28] This circumstance creates a need for gyrokinetic equations that are valid under a more general ordering.

There has been some progress in this direction. Dimits et. al.[29] extended the canonical Hamiltonian gyrokinetic theory[7] to allow for $q\delta\psi/T = O(1)$. Ref. 29 is based on the new small parameter

$$\varepsilon_V \sim \frac{V_\psi}{v_{\text{th}}} \simeq k_\perp \rho_t \frac{q\delta\psi}{T} \ll 1. \quad (5)$$

Here $V_\psi$ is a characteristic drift velocity associated with $\delta\psi$, given by

$$\begin{aligned} V_\psi &= \frac{c}{B}\hat{\boldsymbol{b}} \times \boldsymbol{\nabla}\delta\psi \\ &\approx \frac{c}{B}\hat{\boldsymbol{b}} \times \boldsymbol{\nabla}\delta\phi + v_\parallel \frac{\delta B_\perp}{B}. \end{aligned} \quad (6)$$

The first term is the perturbed $E \times B$ drift velocity, and the second term is the ("magnetic flutter") component of the parallel velocity perpendicular the equilibrium magnetic field. Here, $\delta B_\perp = \left|\boldsymbol{\nabla} \times \left(\hat{\boldsymbol{b}}\delta A_\parallel\right)\right|$ is the magnitude of the perturbed perpendicular magnetic field, and $\hat{\boldsymbol{b}}$ is the unit vector in the direction of the unperturbed (equilibrium) magnetic field. Under this ordering, $\delta\psi$ can have large long-wavelength components with $q\delta\psi/T = O(1)$ and $k_\perp \rho_t \ll 1$ and small short-wavelength components with $q\delta\psi/T \ll 1$ and $k_\perp \rho_t = O(1)$, as well as components of intermediate sizes at intermediate scales. The ordering of Eq. (5) can be called a "drift ordering" in the sense that it is an ordering specifically on the perpendicular velocity, and is consistent with the sum of the $E \times B$ and flutter velocities being limited by the diamagnetic speed, i.e.,

$$\frac{V_\psi}{v_{\text{th}}} \lesssim \frac{\rho_t}{L_p}. \quad (7)$$

The magnetic term in Eq. (6), when inserted into Eq. (5), specifies that the perpendicular velocity associated with magnetic flutter is small compared to the thermal speed or, equivalently, that the angle between the perturbed and unperturbed field lines is small.

Equation (7) is satisfied in observations of strong-gradient regions and, furthermore, the ratio $\rho_t/L_p$ is observed to be small. For example, approximate values of $\rho_t/L_p \sim 1/10$ can be inferred from observations in internal transport barriers in DIII-D[28] and $\rho_t/L_p \sim 1/20$ for internal transport barriers in JT60-U,[28] while measurements for the DIII-D edge[26] indicate $\rho_t/L_p \sim 1/30$.

It can also be argued via a heuristic "mixing-length" argument that Equation (7) is satisfied for typical ion-scale (i.e., ion-gyroradius or longer perpendicular scale) gradient-driven plasma microinstabilities and associated turbulence. Such arguments are typically based on a combination of heuristic reasoning and dimensional analysis, and the exact details vary depending on the particular physics of the instability or turbulent state involved. For a predominantly electrostatic ion-scale instability for which a profile (e.g., density, temperature or pressure) gradient is the primary source of free energy, the fluctuating electron density can be taken to approximately adiabatic (i.e., a Boltzmann response). In many cases, this condition is approximately satisfied even if the electrons themselves have an additional nonadiabatic contribution. Combining quasineutrality and the Boltzmann electron response (for which the small-amplitude form is adequate for estimation), the densities approximately satisfy

$$\delta n_i Z_i \approx \delta n_e \approx n_0 e \delta \phi / T_e, \qquad (8)$$

where $\delta n_i$ is the fluctuation ion number density, $\delta n_e$ is the fluctuating electron number density, $n_0$ is the mean background number density, $-e$ is the electron charge, $q_i = Z_i e$ is the ion charge, and $T_e$ is the electron temperature. If the source of free energy is a profile gradient with spatial scale $L_p$, then the ion density perturbation $\delta n_{i\lambda}$ at a scale $\lambda \leq L_p$ is limited approximately by a value that represents local flattening of the driving profile gradient, i.e.,

$$|\delta n_{i\lambda}| \leq \alpha n_0 \lambda / L_p, \qquad (9)$$

where $\alpha$ is a constant of order 1. Combining Eqs. (8) and (9), gives for the ratio of the $E \times B$ velocity at each spatial scale $\lambda$ to the thermal velocity

$$\frac{V_{E \times B, \lambda}}{v_{\text{th}}} \simeq \frac{\rho_t}{\lambda} \frac{q \delta \phi_\lambda}{T} \lesssim Z_i \alpha \frac{\rho_t}{L_p}. \qquad (10)$$

Note that this expression is independent of $\lambda$.

It has been recognized, for example by Parra and Catto,[10] that the iterative (non-Hamiltonian) derivations are valid under the ordering of Eq. (5). The derivation in Ref. 10 was for the case of a general equilibrium magnetic field and electrostatic perturbations. In a sense, therefore, the key remaining need is for a set of equations that are both electromagnetic, as are those of Ref. 29, and apply to toroidal geometry, as do those of Ref. 10. The possibility exists particularly for edge conditions that the equations will be needed to higher orders in the theory, even under an extended ordering, than are needed for typical core parameters. This favors a Hamiltonian treatment, because at second and higher orders the gyrokinetic system satisfies non-trivial dynamical conservation laws, which are more easily preserved in a Hamiltonian treatment.

An additional property of strong gradient regions, and one which results in a considerable simplification of the resulting gyrokinetic equations, is that $L_p$ is significantly shorter than the scale over which the magnetic equilibrium changes, which for a tokamak is typically the major radius $R$. Example values of $L_p/R$ in internal transport barriers include approximately 1/15 in DIII-D and 1/25 in JT60-U,[28] while in the DIII-D edge values of order 1/50 have been reported.[26] Thus we can use the following "strong-gradient" ordering

$$L_p/R = O(\varepsilon). \qquad (11)$$

The use of this additional ordering has the consequence that instead of Eq. (4), we now have

$$\frac{\rho_t \|\boldsymbol{\nabla} B\|}{B} = O(\varepsilon^2) \qquad (12)$$

so that successive derivatives of the equilibrium magnetic vector potential and magnetic field only arise at even orders in the theory. To second order, terms that couple the gyrophase dependences of the equilibrium magnetic field and of the perturbations are absent. The simpler two-step derivation (discussed earlier in this introduction) can then be used to derive the theory consistently to second order in $\varepsilon$ under the combined orderings of Eqs. (11), (5) and (7)

$$\frac{V_\psi}{v_{\text{th}}} \simeq k_\perp \rho_t \frac{q\delta\psi}{T} \sim \rho/L_p \sim L_p/R \sim \varepsilon \ll 1. \qquad (13)$$

These orderings, which we have argued to be applicable in strong-gradient regions in tokamaks, will therefore form the basis for the present work.

The gyrokinetic equations have also been derived under an ordering that is still more general than that of Eq. (5), and which allows for large $E \times B$ flow velocities (of order the thermal speed).[30-33] If the motion is considered in a frame moving with the local $E \times B$ flow, then it is evident that adiabatic invariance of the magnetic moment can be preserved provided that the difference in the $E \times B$ velocity across a gyro orbit is small compared with the particle's perpendicular (gyration) speed. This translates to a requirement that

$$\frac{V'_{ExB}}{\Omega} \ll 1, \qquad (14)$$

where $V'_{ExB}$ is a measure of the spatial derivatives of (or shearing rate associated with) the $E \times B$ velocity. In the uniform self-consistent theory that results from such an ordering,[33] the equations contain numerous new terms that are challenging to implement numerically, even in the simplified slab electrostatic case addressed in Ref. 33. However, the ExB velocities inferred from measurements in transport barriers[28] and in the edge,[26] are much less than the ion thermal velocity, so that the orderings of Eq. (13) and the resulting much simpler equations are expected to be applicable.

In the present paper we therefore extend the derivation of Ref. 29 to spatially inhomogeneous (including toroidal) equilibrium magnetic fields, under the ordering for the latter of Eq. (11), using noncanonical Hamiltonian methods. As Ref. 29 was already based on the ordering of Eq. (5), the resulting combination of orderings to be used here is exactly that stated in Eq. (13). We develop the gyrocenter Lagrangian consistently to second order in $\varepsilon$.

We anticipate that our equations will be useful as a basis for simulation models for strong gradient regions in magnetized plasmas, e.g., edge and screapeoff layer regions as well as internal transport barriers where the separation between the profile and gyroradius scales is relatively modest, and the significant fluctuations have wavelengths (of the order of 10-100 gyroradii) that may be comparable to the scale of profile variations.

## II. Gyrokinetic equations in the drift ordering

Summarizing the setup and orderings as developed in the introduction, we will consider a plasma in an inhomogeneous time dependent magnetic field $\boldsymbol{B}(\boldsymbol{x},t) = \boldsymbol{B}_0(\boldsymbol{x}) + \delta\boldsymbol{B}(\boldsymbol{x},t)$, where $\boldsymbol{B}_0(\boldsymbol{x}) = B_0(\boldsymbol{x})\hat{\boldsymbol{b}}_0(\boldsymbol{x})$ is an equilibrium magnetic field, and we define $B_0(\boldsymbol{x}) = |\boldsymbol{B}_0(\boldsymbol{x})|$, so that $\hat{\boldsymbol{b}}_0(\boldsymbol{x})$ is the unit vector in the direction of $\boldsymbol{B}_0$. We allow for electromagnetic perturbations consisting of a perturbed electrostatic potential $\delta\phi = \delta\phi(\boldsymbol{x},t)$ and a perturbed magnetic field in a low-$\beta$ ordering, where $\beta$ is the ratio of plasma to magnetic pressure, $\delta\boldsymbol{B}(\boldsymbol{x},t) = \boldsymbol{\nabla} \times \left[\delta A_\parallel(\boldsymbol{x},t)\hat{\boldsymbol{b}}_0(\boldsymbol{x})\right]$. The generalized potential $\delta\psi = \delta\phi - v_\parallel \delta A_\parallel$ is taken to satisfy $\hat{\boldsymbol{b}}_0 \times \boldsymbol{\nabla}\delta\psi/(\Omega v_{\text{th}}) \ll 1$. Here, $\Omega$ denotes the gyrofrequency associated with the equilibrium magnetic field.

Consistent with Eq. (13) and with having the equilibrium magnetic field at zero order, we take

$$qA_0/(Mcv_{\text{th}}) = O(\varepsilon^{-2}).$$

Because the perturbed ($E \times B$ and magnetic-flutter) velocities are small, they will not have a zero-order contribution to the symplectic (noncanonical) terms in the particle Lagrangian. We can therefore begin with the unperturbed guiding-center phase-space variables

$$Z = (\boldsymbol{R}, U_\parallel, \mu, \theta),$$

where $\boldsymbol{R}$ is the guiding center position, $U_\parallel$ is the parallel velocity, $\mu$ is the magnetic moment, and $\theta$ is the gyrophase angle. The (guiding-center) Jacobian of the transformation from $(\boldsymbol{x}, \boldsymbol{v})$ to $Z$ is[9]

$$\Omega^* \equiv \left\| \frac{\partial(\boldsymbol{x}, \boldsymbol{v})}{\partial Z} \right\| = \Omega + U_{\|} \hat{\boldsymbol{b}}_0 \cdot \boldsymbol{\nabla} \times \hat{\boldsymbol{b}}_0. \tag{15}$$

The guiding-center variables $Z$ evolve according to the Euler-Lagrange equations with the standard second order guiding-center phase-space Lagrangian for the motion of a charged particle in a static inhomogeneous magnetic field:[2]

$$\mathcal{L}(Z, \dot{Z}, t) = \left[ \boldsymbol{A}_0(\boldsymbol{R}) + U_{\|} \hat{\boldsymbol{b}}_0(\boldsymbol{R}) - \mu \boldsymbol{G} \right] \cdot \dot{\boldsymbol{R}} - \mu \dot{\theta} - \left[ \frac{U_{\|}^2}{2} + \mu \Omega(\boldsymbol{R}) + \frac{1}{2} \mu U_{\|} \hat{\boldsymbol{b}}_0 \cdot \boldsymbol{\nabla} \times \hat{\boldsymbol{b}}_0 \right], \tag{16}$$

where $\boldsymbol{G} = (\boldsymbol{\nabla} \boldsymbol{e}_1) \cdot \boldsymbol{e}_2$ is Littlejohn's gyro-gauge vector field,[2] and $\boldsymbol{e}_1$, $\boldsymbol{e}_2$ and $\hat{\boldsymbol{b}}_0$ form a ($\boldsymbol{R}$-dependent) right orthonormal set of unit vectors. The $\boldsymbol{G}$ term ensures that the gyroradius vector is independent of the choice of orientation ("gyrogauge") of the two, generally spatially dependent, perpendicular unit vectors $\boldsymbol{e}_1$ and $\boldsymbol{e}_2$, while the $\frac{1}{2} \mu U_{\|} \hat{\boldsymbol{b}}_0 \cdot \boldsymbol{\nabla} \times \hat{\boldsymbol{b}}_0$ term leads to the Baños drift[34] and a torsional contribution to $\dot{\theta}$. It is noted in the discussion surrounding Eq. (147) of Ref. 26 that this last term is consistent with the results in that reference. We normalize energies and the Lagrangian to the thermal energy, ($\mathcal{L} \sim T$), velocities (e.g., $U_{\|}$) to the mean thermal velocity $v_{\text{th}}$, momenta to $M v_{\text{th}}$, magnetic potentials ($\boldsymbol{A}_0$, $\delta \boldsymbol{A}$) to the quantity $M v_{\text{th}} c / q$. Electrostatic potentials ($\delta \phi$) will be normalized to $T/q$. It can easily be checked that the Euler-Lagrange equations applied to Eq. (16) give the standard guiding-center equations of motion with parallel streaming, and the $\boldsymbol{\nabla} B$ and curvature drifts.

To the Lagrangian of Eq. (16), we add the perturbed Lagrangian

$$\delta \mathcal{L} = \delta A_{\|}(\boldsymbol{x}, t) \hat{\boldsymbol{b}}_0(\boldsymbol{x}) \cdot \dot{\boldsymbol{x}} - \delta \phi(\boldsymbol{x}, t). \tag{17}$$

There are two standard choices for the perturbed guiding-center phase-space variables. In the "symplectic representation," $U_{\|}$ is used as the parallel momentum variable, while in the "canonical representation," a perturbed canonical momentum $p_{\|} = U_{\|} + \delta A_{\|}$ is used. Thus, in the canonical representation the perturbed guiding-center coordinates and Lagrangian are $Z = (\boldsymbol{R}, p_{\|}, \mu, \theta)$ and

$$\mathcal{L}(Z, \dot{Z}, t) = \left[ \boldsymbol{A}_0(\boldsymbol{R}) + p_{\|} \hat{\boldsymbol{b}}_0(\boldsymbol{R}) - \mu \boldsymbol{G} \right] \cdot \dot{\boldsymbol{R}} - \mu \dot{\theta}$$
$$- \left\{ \frac{[p_{\|} - \delta A_{\|}(\boldsymbol{R} + \boldsymbol{\rho}, t)]^2}{2} + \mu \Omega(\boldsymbol{R}) + \frac{1}{2} \mu (p_{\|} - \delta A_{\|}) \hat{\boldsymbol{b}}_0 \cdot \boldsymbol{\nabla} \times \hat{\boldsymbol{b}}_0 + \delta \phi(\boldsymbol{R} + \boldsymbol{\rho}, t) \right\},$$

while in the symplectic representation they are $Z = (\boldsymbol{R}, U_{\|}, \mu, \theta)$ and

$$\mathcal{L}(Z,\dot{Z},t) = \{A_0(R) + [U_\parallel + \delta A_\parallel(R+\rho,t)]\hat{b}_0(R) - \mu G\}\cdot \dot{R} - \mu\dot{\theta}$$
$$- \left[\frac{U_\parallel^2}{2} + \mu\Omega(R) + \frac{1}{2}\mu U_\parallel \hat{b}_0 \cdot \nabla\times\hat{b}_0 + \delta\phi(R+\rho,t)\right].$$

For the remainder of this paper, we proceed in the canonical representation, as this is somewhat simpler than the symplectic representation. The derivation of the gyrocenter Lagrangian for the symplectic case follows in a similar manner, but the results are less transparent in several ways. First, the equations of motion are more challenging to obtain directly (but can be obtained perturbatively) because of the presence of several noncanonical (symplectic) terms in the Lagrangian at second order. Secondly, casting the conserved system energy into a positive-definite quadratic form becomes more challenging.

Proceeding in the canonical representation, separate $\delta\phi$ and $\delta A_\parallel$ as

$$\delta\phi(R+\rho,t) = \delta\bar{\phi}(R,t) + \delta\tilde{\phi}(R,\mu,\theta,t)$$

and

$$\delta A_\parallel(R+\rho,t) = \delta\bar{A}_\parallel(R,t) + \delta\tilde{A}_\parallel(R,\mu,\theta,t),$$

where $\delta\bar{\phi}(R,\mu,t)$ and $\delta\bar{A}_\parallel(R,\mu,t)$ are gyrophase-independent approximations to $\delta\phi$ and $\delta A_\parallel$. A convenient choice for these, and one which will simplify the subsequent derivations is

$$\delta\bar{\psi}(R,t) = \langle\delta\psi\rangle$$
$$= \frac{1}{2\pi}\oint d\theta\,\delta\psi(R+\rho\hat{\rho}(\theta),t)$$

where $\rho = \sqrt{\dfrac{2\mu}{\Omega(R)}}$, and $\hat{\rho}(\theta)$ is a unit vector perpendicular to $\hat{b}_0(R)$ and which subtends an angle $\theta$ with respect to a fixed plane containing $\hat{b}_0(R)$. Then

$$\delta\tilde{\phi} \equiv \phi(R+\rho,t) - \bar{\phi}(R,\mu,t) \approx \rho\cdot\nabla\bar{\phi} = O\left(V_{E\times B}/v_{\text{th}}\right) = O(\varepsilon_V), \quad (18)$$

where $\varepsilon_V$ is the small ordering parameter, as given in the ordering relation Eq. (5). Similarly,

$$\delta\tilde{A}_\parallel \equiv \delta A_\parallel(R+\rho,t) - \delta\bar{A}_\parallel(R,\mu,t) \approx \rho\cdot\nabla\delta\bar{A}_\parallel = O(\delta B_\perp/B_0) = O(\varepsilon_V). \quad (19)$$

After the transformation to the perturbed guiding-center variables and the above separation of $\delta\tilde{\phi}$ and $\delta\tilde{A}_\parallel$, the resulting Lagrangian is

$$\mathcal{L} = \langle \mathbf{A}_0 \rangle \cdot \dot{\mathbf{R}} \qquad \qquad \ldots\ldots O(\varepsilon^{-2})$$

$$+ p_\| \hat{\mathbf{b}}_0 \cdot \dot{\mathbf{R}} - \mu\dot{\theta} - \left[\frac{1}{2}\left(p_\| - \delta\overline{A}_\|\right)^2 + \mu\Omega + \delta\overline{\phi}\right] \qquad \ldots\ldots O(\varepsilon^0)$$

$$+ \left[\delta\tilde{\phi} - \left(p_\| - \delta\overline{A}_\|\right)\delta\tilde{A}_\|\right] \qquad \ldots\ldots O(\varepsilon^1) \qquad (20)$$

$$- \mu \mathbf{G} \cdot \dot{\mathbf{R}} - \left[\frac{1}{2}\left(\delta\tilde{A}_\|\right)^2 + \frac{1}{2}\mu\left(p_\| - \delta\overline{A}_\|\right)\hat{\mathbf{b}}_0 \cdot \boldsymbol{\nabla}\times\hat{\mathbf{b}}_0\right], \qquad \ldots\ldots O(\varepsilon^2)$$

where the terms in Eq.(20) are separated into terms formally of order $\varepsilon^{-2}$ through $\varepsilon^2$ in the expansion parameter in the collected set of orderings Eq. (13). The key result of the separation in Eqs. (18) and (19) is that the only gyrophase dependent terms in Eq. (20) are at orders $\varepsilon^1$ and higher.

A Lie-transform perturbative treatment is applied to transform the phase-space coordinates $Z \to \overline{Z} = \left(\overline{\mathbf{R}}, \overline{U}_\|, \overline{\mu}, \overline{\theta}\right)$ to eliminate the gyrophase dependence in the Lagrangian of Eq. (20) using $\delta\tilde{\phi} \sim \varepsilon \ll 1$. The derivation is similar to the standard noncanonical Hamiltonian Lie transform perturbation theory,[8,9] but also has some differences. The coordinate transformation is represented as an operator $\mathsf{T}(\varepsilon)$, the action of which on the coordinates, distribution function, and Poincare-Cartan one-form $\gamma$ (defined by $\gamma = \mathcal{L}\, dt$, where $\mathcal{L}$ is the Lagrangian), is formally given by

$$Z \to \overline{Z} = \left(\overline{\mathbf{R}}, \overline{U}_\|, \overline{\mu}, \overline{\theta}\right) = \mathsf{T}(\varepsilon) Z,$$
$$f(Z) = \overline{f}(\overline{Z}) = \mathsf{T}\,\overline{f}(Z), \qquad (21)$$
$$\Gamma \equiv \overline{\gamma} = \mathsf{T}^{-1}\gamma + dS,$$

where $S$ is a gauge function. $\mathsf{T}(\varepsilon)$ is further represented as a product of operators, each of which is formally an exponential of Lie derivative operators and which act only at successively higher orders

$$\mathsf{T} = \ldots\ldots \mathsf{T}_3 \mathsf{T}_2 \mathsf{T}_1,$$
$$\mathsf{T}_n(\varepsilon) = \exp(\varepsilon^n L_n). \qquad (22)$$

Here the $L_n$'s are Lie derivatives (not to be confused with the Lagrangian $\mathcal{L}$). The operator $\mathsf{T}$ in Eq. (22) and the one-form $\Gamma$ in Eqs. (21) are expanded in powers of $\varepsilon$. A difference with most other derivations is that operators involving spatial derivatives are assigned different orders depending on the quantity upon which they operate.[20] Spatial derivatives acting on a quantity at a given order in the Lagrangian of Eq. (20) may demote that quantity zero, one or two orders. The key steps in the evaluation of $\Gamma$ are presented in the Appendix.

The resulting phase-space Lagrangian for the gyrocenter motion, in the canonical representation for the magnetic perturbations, and up to second order is

$$\mathcal{L} = \left[\langle \boldsymbol{A}_0 \rangle + p_\parallel \hat{\boldsymbol{b}}_0 - \mu \boldsymbol{G}\right] \cdot \dot{\boldsymbol{R}} - \mu \dot{\theta} - H, \tag{23}$$

where $H$ is the Hamiltonian

$$\begin{aligned} H = &\frac{1}{2}\left\langle \left(p_\parallel - \delta A_\parallel\right)^2 \right\rangle + \mu \Omega + \langle \delta\phi \rangle \\ &- \frac{1}{2\Omega}\left\langle \boldsymbol{\nabla}\left(\tilde{\Psi}/\Omega_0\right) \times \hat{\boldsymbol{b}}_0 \cdot \boldsymbol{\nabla}\tilde{\psi} \right\rangle - \frac{1}{2\Omega}\frac{\partial}{\partial \mu}\langle \tilde{\psi}^2 \rangle \\ &+ \frac{1}{2}\mu\left(p_\parallel - \delta\overline{A}_\parallel\right)\hat{\boldsymbol{b}}_0 \cdot \boldsymbol{\nabla} \times \hat{\boldsymbol{b}}_0, \end{aligned} \tag{24}$$

$$\begin{aligned} \tilde{\psi} &= \delta\tilde{\phi} - \left(p_\parallel - \delta\overline{A}_\parallel\right)\delta\tilde{A}_\parallel, \\ \tilde{\Psi} &= \Psi - \langle \Psi \rangle, \\ \Psi &= \int^\theta \tilde{\psi}\, d\theta, \end{aligned}$$

and

$$\langle \Psi \rangle = \frac{1}{2\pi}\oint \tilde{\Psi}\, d\theta.$$

This is essentially the same result as has been obtained in the standard ordering. For example, Eqs. (23) and (24) are the same as the sum of Eqs. (148), (172) and (173) of Ref. 9, with the exception that we have retained one higher order in the equilibrium magnetic-fields terms, and because under our ordering the identification of the terms with orders in $\varepsilon$ is different. These additional equilibrium terms are retained so that the resulting Lagrangian is consistent to second order in our ordering. The key difference with previous work is that we have provided a justification for the Lagrangian of Eqs. (23) and (24) in the combination of the strong-gradient orderings of Eqs. (13), which is relevant to the edge region and internal transport barriers in MFE devices.

The equations of motion follow from the Euler-Lagrange equations applied with the Lagrangian of Eq. (23). These are

$$\dot{\boldsymbol{R}} = \frac{\hat{\boldsymbol{b}}_0}{B_\parallel^*} \times \boldsymbol{\nabla}_R H + \frac{\boldsymbol{B}^*}{B_\parallel^*}\left(p_\parallel - \langle \delta A_\parallel \rangle\right),$$

$$\dot{p}_\parallel = -\frac{\boldsymbol{B}^*}{B_\parallel^*} \cdot \boldsymbol{\nabla}_R H,$$

$$\dot{\mu} = 0,$$

$$\dot{\theta} = -\frac{\partial H}{\partial \mu}.$$

(25)

where

$$\boldsymbol{B}^* \equiv \boldsymbol{B}_0 + p_\parallel \boldsymbol{\nabla} \times \hat{\boldsymbol{b}}_0 - \mu \boldsymbol{\nabla} \times \boldsymbol{G},$$

$$B_\parallel^* \equiv \boldsymbol{B}^* \cdot \hat{\boldsymbol{b}}_0.$$

Equations (25) yield familiar results, including the $\nabla B$, curvature, gyroaveraged $E \times B$ and Baños drifts, along with the corrections to the $E \times B$ drift due to the terms in the Hamiltonian that are quadratic in the perturbations.

The Vlasov equation for the evolution of the gyrocenter distribution function, neglecting collisions, can be obtained in the standard way using the fact that the absence of dependence of Lagrangian on $\theta$ decouples the gyrophase dependent and independent parts of the Vlasov equation.[9] The gyrophase averaged gyrocenter distribution function $F_i$ then satisfies

$$\frac{\partial F_i}{\partial t} + \dot{\boldsymbol{R}} \cdot \boldsymbol{\nabla}_R F_i + \dot{U}_\parallel \frac{\partial F_i}{\partial U_\parallel} = 0. \qquad (26)$$

In a self-consistent model, the gyrokinetic species described by the above equations will contribute to the field equations through its density $n_i$ and current $\boldsymbol{J}_i$. These are obtained from $F_i$ via the "pullback transformation" $\bar{f}(\bar{Z}) = \mathsf{T}\bar{f}(Z)$ and integration over velocity space. Using Eqs. (22), and (35) (given in the appendix), and keeping terms in $\mathsf{T}$ up to first order, the results are for the density

$$n_i \approx q \int \Omega^* \; d^6 Z \; \delta(\boldsymbol{R} + \boldsymbol{\rho} - \boldsymbol{x}) \left[ F_i(\boldsymbol{R}, \mu) + \frac{1}{\Omega} \tilde{\psi} \frac{\partial F_i}{\partial \mu} + \frac{1}{\Omega} \boldsymbol{\nabla}_{R\perp} \left(\frac{\tilde{\Psi}}{\Omega}\right) \cdot \hat{\boldsymbol{b}}_0 \times \boldsymbol{\nabla}_{R\perp} F_i \right], \qquad (27)$$

and for the parallel ion current

$$\delta J_{\parallel i} \approx q \int \Omega^* \; d^6 Z \; \delta(\boldsymbol{R} + \boldsymbol{\rho} - \boldsymbol{x}) \, p_\parallel \left( F_i + \frac{1}{\Omega} \tilde{\psi} \frac{\partial F_i}{\partial \mu} + \frac{1}{\Omega} \boldsymbol{\nabla}_{R\perp} \left(\frac{\tilde{\Psi}}{\Omega}\right) \cdot \hat{\boldsymbol{b}}_0 \times \boldsymbol{\nabla}_{R\perp} F_i \right)$$
$$- q n_i \delta A_\parallel. \qquad (28)$$

Note that the Jacobian used in the integration is the guiding-center Jacobian of Eq. (15).

Inserting Eq. (27) and (28) respectively into the Poisson equation and Ampere's law, for the case of a single gyrokinetic ion species, and neutralizing electrons (with the electron charge denoted as $-e$) gives

$$\nabla^2 \phi = 4\pi e n_e$$
$$- 4\pi q \int \Omega^* d^6 Z \; \delta(\boldsymbol{R} + \boldsymbol{\rho} - \boldsymbol{x}) \left[ F_i + \frac{1}{\Omega} \tilde{\psi} \frac{\partial F_i}{\partial \mu} + \frac{1}{\Omega} \boldsymbol{\nabla}_{R\perp} \left( \frac{\tilde{\Psi}}{\Omega} \right) \cdot \hat{\boldsymbol{b}_0} \times \boldsymbol{\nabla}_{R\perp} F_i \right] \quad (29)$$

and

$$\left( \nabla^2 / c^2 \right) A_{1\parallel} = \left( 4\pi e / m_e \right) \int dV \; F_e p_\parallel - \left( 4\pi q / M \right) \times$$
$$\left\{ \int \Omega^* \; d^6 Z \; \delta(\boldsymbol{R} + \boldsymbol{\rho} - \boldsymbol{x}) \, p_\parallel \left[ F_i + \frac{1}{\Omega} \tilde{\psi} \frac{\partial F_i}{\partial \mu} + \frac{1}{\Omega} \boldsymbol{\nabla}_{R\perp} \left( \frac{\tilde{\Psi}}{\Omega} \right) \cdot \hat{\boldsymbol{b}_0} \times \boldsymbol{\nabla}_{R\perp} F_i \right] - n_i \delta A_\parallel \right\}, \quad (30)$$

where $m_e$ is the electron mass. Equations (23) and the associated Euler-Lagrange equations, along with Eqs. (26), (29), and (30) constitute a closed gyrokinetic Vlasov-Maxwell system of equations valid for strong-gradient regions in a magnetized plasma, under the orderings of Eq. (13).

An important result of the variational formulation[9,21] ("energetic" and momentum-conservation "consistency") is that in order for the conservation laws to hold, if the Lagrangian that generates the equations of motion is calculated to second order as is the case for Eqs. (23) and (24), then terms up to first order only should be kept in the pullback transformation used to calculate the density and current that are used in Poisson's equation and Ampere's law, as is the case for Eqs. (27)-(30). The resulting conserved system energy for the gyrokinetic system given by Eqs. (23), (26), (29) and (30) is

$$E_{\text{tot}} = \int \Omega^* \; d^6 Z \; F_i \left( H - q \langle \mathsf{T}_1^{-1} \delta \phi \rangle \right) + E_e + \int \frac{d^3 x}{8\pi} \left( |\boldsymbol{\nabla} \delta \phi|^2 + |\boldsymbol{B}|^2 \right),$$

where $E_e$ is the electron energy.

### III. Summary

We have derived the (low-$\beta$) toroidal electromagnetic gyrokinetic equations systematically to second order in a an extended ordering which allows for large perturbation amplitudes and is appropriate for strong-gradient regions in MFE plasmas, such as the edge and scrapeoff-layer and internal transport barriers. While the resulting equations are similar to those already published, the present work is valuable because it provides a theoretical basis for the application of these equations in strong-gradient regions in magnetized plasmas, for example in the edge region and internal transport barriers in a tokamak. In particular, our ordering and derivation show that the equations are valid in such regions, while previous Hamiltonian derivations used an ordering that precluded large perturbations. Also, while the ordering used in the present paper is not quite as general as some of the large-flow orderings,[30-33] the resulting equations are much simpler than those that result from a uniform self-consisstent application of the large-flow ordering for a self-consistent system.[33]

Another simplification in the present work is the use of a strong-gradient ordering of the plasma profiles relative to the magnetic-field inhomogeneities, which results in equations valid consistently to second order without the many additional terms that must be kept in the standard gyrokinetic orderings.[11] Thus, the simplicity of the more familiar second-order equations is maintained,[9] but with a stronger justification. Because the equations are valid to second order and come from a Hamiltonian formalism, energy and momentum conservation relations that are satisfied exactly follow directly by application of Noether's theorem in a variational formulation.[9,21] This is an advantage over non Hamiltonian derivations, for which the conservation relations that result are typically approximate rather than exact.

While our derivation is collisionless, the ordering and particular results for the Lie transformations of the coordinates and distribution function form the basis for the numerical implementation and evaluation of Coulomb and ion-neutral collisions based on the gyrophase averaged gyrocenter distribution function. Such implementations are valid provided that the collisional coupling between the gyrophase-dependent and independent parts of the distribution function can be ordered as a small term.

We anticipate that our equations will be useful as a basis for simulation models edge and screapeoff layer regions as well as internal transport barriers in tokamaks, where the separation between the profile and gyroradius scales is relatively modest, and the significant fluctuations have wavelengths (of the order of 10-100 gyroradii) that may be comparable to the scale of profile variations.

## Appendix: Derivation of the perturbed Lie-transform generators and gyrocenter Lagrangian

Here, we show the key steps in the calculations leading from Eqs. (20)-(22) to Eq. (23).

The action of Lie derivatives $L_n$ on a scalar and one form are

$$L_n \Lambda = g_n^\beta \frac{\partial \Lambda}{\partial Z^\beta},$$
$$\left(L_n \gamma\right)_\alpha = g_n^\beta \left(\frac{\partial \gamma_\alpha}{\partial Z^\beta} - \frac{\partial \gamma_\beta}{\partial Z^\alpha}\right). \tag{31}$$

Spatial derivatives acting on a quantity at a given order in the Lagrangian of Eq. (20) may demote that quantity zero, one or two orders, so that operators involving these spatial derivatives are assigned different orders depending on the quantity upon which they operate.[20] We therefore, write

$$L_n \gamma = (L_n \gamma)_a + \varepsilon (L_n \gamma)_b + \varepsilon^2 (L_n \gamma)_c$$

The main results needed from the perturbation theory to obtain the Lagrangian up to second order are

$$\begin{aligned}
\Gamma_{-2,-1,0} &= \gamma_{-2,-1,0} + dS_{-2,-1,0}, \\
\Gamma_1 &= \gamma_1 - (L_1 \gamma_0)_a - (L_1 \gamma_{-2})_c + dS_1, \\
\Gamma_2 &= \langle \gamma_2 \rangle - \frac{1}{2} \langle (L_1 \gamma_1)_a \rangle.
\end{aligned} \quad (32)$$

For orders $\varepsilon^{-2}$, $\varepsilon^{-1}$, and $\varepsilon^0$ in Eq. (32), we can choose

$$\begin{aligned}
\Gamma_{-2,-1,0} &= \gamma_{-2,-1,0}, \\
dS_{-2,-1,0} &= 0.
\end{aligned}$$

At first order, we have

$$0 = \langle \gamma_1 \rangle = \Gamma_1 = \gamma_1 - (L_1 \gamma_0)_a - (L_1 \gamma_{-2})_c + dS_1,$$

which yields

$$\begin{aligned}
\boldsymbol{g}_1^{\perp} &= -\frac{1}{\Omega_0} \hat{\boldsymbol{b}}_0 \times \boldsymbol{\nabla} S_1, \\
g_1^{\parallel} &= -\frac{\partial S_1}{\partial p_{\parallel}}, \\
g_1^{p_{\parallel}} &= \nabla_{\parallel} S_1, \\
g_1^{\mu} &= -\frac{\partial S_1}{\partial \theta}, \\
g_1^{\theta} &= \frac{\partial S_1}{\partial \mu}, \\
0 &= \frac{\partial S_1}{\partial t} + \Omega g_1^{\mu} + \left( p_{\parallel} - \delta \overline{A}_{\parallel} \right) g_1^{p_{\parallel}} - \tilde{\psi},
\end{aligned} \quad (33)$$

where

$$\tilde{\psi} \equiv \delta \tilde{\phi} - \left( p_{\parallel} - \delta \overline{A}_{\parallel} \right) \delta \tilde{A}_{\parallel}.$$

these, in turn, result in

$$\frac{\partial S_1}{\partial t} + \left(p_\| - \delta\overline{A}_\|\right)\nabla_\| S_1 - \Omega\frac{\partial S_1}{\partial\theta} = \tilde{\psi}. \qquad (34)$$

The first-order gauge function $S_1$ is needed to calculate the first-order generating functions and the expression for the second-order parts of the Lagrangian. In the solution of Eq. (34) for $S_1$ we use $\partial S_1/\partial t \sim \varepsilon\Omega\partial_\theta S_1$, from which follows

$$\left(\frac{dS_1}{dt}\right)_{\text{slow}} - \Omega\frac{\partial S_1}{\partial\theta} = \tilde{\psi},$$

$$S_1 \approx -\tilde{\Psi}/\Omega,$$

$$\tilde{\Psi} \equiv \Psi_i - \overline{\Psi}_i,$$

$$\Psi_i = \int_{\theta_0}^{\theta} d\theta\,\delta\tilde{\psi}.$$

Inserting this result into Eqs. (33) gives

$$\boldsymbol{g}_1^\perp = \frac{1}{\Omega}\hat{\boldsymbol{b}}_0 \times \boldsymbol{\nabla}\left(\tilde{\Psi}/\Omega\right),$$

$$g_1^\| = -\Delta\tilde{A}_\|/\Omega,$$

$$g_1^{p_\|} = -\nabla_\|\left(\tilde{\Psi}/\Omega\right), \qquad (35)$$

$$g_1^\mu = \left(\tilde{\psi}/\Omega\right),$$

$$g_1^\theta = -\frac{1}{\Omega}\frac{\partial\tilde{\Psi}}{\partial\mu},$$

where

$$\tilde{\Psi} \equiv \tilde{\Phi} - \left(p_\| - \delta\overline{A}_\|\right)\Delta\tilde{A}_\|,$$

$$\tilde{\Phi} \equiv \Phi_i - \overline{\Phi}_i,$$

$$\Phi_i = \int_{\theta_0}^{\theta} d\theta\,\delta\tilde{\phi},$$

$$\Delta\tilde{A}_\| = \Delta A_{\|i} - \Delta\overline{A}_{\|i},$$

$$\Delta A_{\|i} = \int_{\theta_0}^{\theta} d\theta\,\delta\tilde{A}_\|.$$

The resulting Poincare-Cartan one-form (phase-space Lagrangian) for the gyrocenter motion, in the canonical representation for the magnetic perturbations and up to second order, is

$$\begin{aligned}
\Gamma = & \langle \mathbf{A}_0 \rangle \cdot d\mathbf{R} - \mu \mathbf{G} \cdot d\mathbf{R} \\
& + p_\parallel \hat{\mathbf{b}}_0 \cdot d\mathbf{R} - \mu\, d\theta - \left[\frac{1}{2}\left\langle \left(p_\parallel - \delta A_\parallel\right)^2 \right\rangle + \mu \Omega + \langle \delta\phi \rangle\right] dt \\
& + \frac{1}{2}\left[\frac{1}{\Omega}\frac{\partial \langle \tilde{\psi}^2 \rangle}{\partial \mu} + \frac{1}{\Omega}\left\langle \hat{\mathbf{b}}_0 \times \boldsymbol{\nabla}\left(\frac{\tilde{\Psi}}{\Omega}\right) \cdot \boldsymbol{\nabla}\tilde{\psi} \right\rangle \right. \\
& \left. - \mu\left(p_\parallel - \delta\bar{A}_\parallel\right)\hat{\mathbf{b}}_0 \cdot \boldsymbol{\nabla} \times \hat{\mathbf{b}}_0 + \nabla_\parallel\left(\frac{1}{\Omega}\langle \delta\tilde{A}_\parallel \tilde{\Psi} \rangle\right)\right] dt
\end{aligned} \qquad (36)$$

The last ($\nabla_\parallel$) term in the square brackets in Eq. (36) can be neglected because it is third order in $\varepsilon$. This yields the result given in Eq. (23).

**Acknowledgments**

The author wishes to acknowledge useful discussions with and suggestions from I. Calvo, J. Candy, P. Catto, B. Cohen, R. Cohen, M. Dorf, D. Ernst, G. Hammett, J. Krommes, F. Parra, T. Rognlien, R. Waltz and X. Q. Xu. This work was performed for U.S. DOE by LLNL under Contract DE-AC52-07NA27344, and is a part of the ESL.